\documentclass[pre,aps,10pt,twocolumn,floatfix,superscriptaddress]{revtex4-1}

\usepackage[utf8]{inputenc}
\usepackage[nice]{nicefrac}
\usepackage{graphicx,color}
\usepackage{amsmath,amssymb,bm,bbm}
\usepackage{amsmath}
\usepackage{braket}
\usepackage{xcolor}
\usepackage{float}

\usepackage[nice]{nicefrac}

\usepackage{physics} 

\usepackage[T1]{fontenc}
\usepackage[utf8]{inputenc}

\DeclareUnicodeCharacter{2212}{~}

\usepackage[plainpages=false,pdfpagelabels,colorlinks=true,linkcolor=red,urlcolor=blue,citecolor=blue,pdftitle={},pdfauthor={},pdfdisplaydoctitle=true,pdfduplex=DuplexFlipLongEdge]{hyperref}

\definecolor{max}{HTML}{03A678}
\definecolor{rafal}{HTML}{161226}

\definecolor{darkgreen}{rgb}{0,0.60,.2}
\definecolor{darkblue}{rgb}{0.1,0.3,1}

\usepackage{physics} 
\usepackage[normalem]{ulem} 
\usepackage[T1]{fontenc}
\usepackage[utf8]{inputenc}

\usepackage{amsthm}
\usepackage{amssymb}

\date{\today}

\begin{document}

\title{Eigenstate entanglement entropy in the integrable spin-\texorpdfstring{$\frac{1}{2}$}{1/2} $XYZ$ model}

\author{R. Świętek}
\affiliation{Department of Theoretical Physics, J. Stefan Institute, SI-1000 Ljubljana, Slovenia\looseness=-1}
\affiliation{Department of Physics, Faculty of Mathematics and Physics, University of Ljubljana, SI-1000 Ljubljana, Slovenia\looseness=-1}
\author{M. Kliczkowski}
\affiliation{Institute of Theoretical Physics, Faculty of Fundamental Problems of Technology, Wrocław University of Science and Technology, 50-370 Wrocław, Poland\looseness=-3}
\author{L. Vidmar}
\affiliation{Department of Theoretical Physics, J. Stefan Institute, SI-1000 Ljubljana, Slovenia\looseness=-1}
\affiliation{Department of Physics, Faculty of Mathematics and Physics, University of Ljubljana, SI-1000 Ljubljana, Slovenia\looseness=-1}
\author{M. Rigol}
\affiliation{Department of Physics, The Pennsylvania State University, University Park, Pennsylvania 16802, USA}

\begin{abstract}
We study the average and the standard deviation of the entanglement entropy of highly excited eigenstates of the integrable interacting spin-$\frac{1}{2}$ $XYZ$ chain away from and at special lines with U(1) symmetry and supersymmetry. We universally find that the average eigenstate entanglement entropy exhibits a volume-law coefficient that is smaller than that of quantum-chaotic interacting models. At the supersymmetric point, we resolve the effect that degeneracies have on the computed averages. We further find that the normalized standard deviation of the eigenstate entanglement entropy decays polynomially with increasing system size, which we contrast with the exponential decay in quantum-chaotic interacting models. Our results provide state-of-the art numerical evidence that integrability in spin-$\frac{1}{2}$ chains reduces the average, and increases the standard deviation, of the entanglement entropy of highly excited energy eigenstates when compared with those in quantum-chaotic interacting models.
\end{abstract}

\maketitle
\section{Introduction}

In the study of many-body quantum systems, entanglement has been a landmark in identifying novel quantum phases~\cite{Kitaev2005, amico_vlatko_2008, calabrese_cardy_09, eisert2010colloquium, JiangNature2012} and in understanding thermalization under unitary quantum dynamics~\cite{polkovnikov11, dalessio_kafri_16}. Being the result of nonclassical correlations between subsystems in an overall pure state makes entanglement and its measures specially appealing to gain insights on quantum effects. Recently, motivated by questions in the field of quantum thermalization, the scaling of the entanglement entropy of highly excited eigenstates of local Hamiltonians has attracted particular attention. As reference stands Page's finding that the entanglement entropy of typical (Haar-random) pure states in the Hilbert space exhibits a leading volume-law term that is maximal~\cite{page_1993b}. Multiple studies have reported compelling evidence that the average entanglement entropy of highly excited eigenstates of quantum-chaotic interacting Hamiltonians exhibit the same leading volume-law term~\cite{vidmar_rigol_17, garrison_grover_18} (see also Refs.~\cite{beugeling_andreanov_15, yang_chamon_15, dymarsky_lashkari_18} and Ref.~\cite{bianchi_hackl_22} for a pedagogical review). The deviations from Page's result have been found to occur at the level of the $O(1)$ corrections~\cite{huang_21, Haque_Khaymovich_2022, Kliczkowski_Swietek2023, Rodriguez_Khemani_2023}, which we argued in Ref.~\cite{Kliczkowski_Swietek2023} depend on the model Hamiltonian.

Highly-excited eigenstates of integrable interacting Hamiltonians, such as the spin-$\frac{1}{2}$ $XXZ$ chain~\cite{leblond_mallayya_19} and the spin-$\frac{1}{2}$ Heisenberg ($XXX$) chain~\cite{patil2023}, on the other hand have been found to exhibit an average entanglement entropy whose leading volume-law term is not maximal. It is actually close to the one of translationally invariant quadratic models~\cite{vidmar_hackl_17, vidmar_hackl_18, hackl_vidmar_19}, and of quantum-chaotic quadratic models~\cite{lydzba_rigol_20, lydzba_rigol_21, ulcakar_vidmar_22, suntajs_prosen_23} and typical (Haar-random) Gaussian pure states in the Hilbert space~\cite{bianchi_hackl_21, bianchi_hackl_22}. Since the previously mentioned studies of the average eigenstate entanglement entropy of integrable interacting Hamiltonians were carried out in the presence of U(1)~\cite{leblond_mallayya_19} and $SU(2)$~\cite{patil2023} symmetry, our goal in this work is to explore what happens in the absence of those symmetries as well as when the model exhibits supersymmetry (in short, SUSY). For this purpose, we use full exact diagonalization to study the spin-$\frac{1}{2}$ $XYZ$ chain, which is integrable and exhibits SUSY at special lines in its parameter space. We contrast our results with those at a U(1) point. 

The other main question we address in this work is how the normalized standard deviation of the eigenstate entanglement entropy decays with increasing system size. We study this quantity both at integrability and away from integrability. We show that, while it decays exponentially in the quantum-chaotic regime, at integrability it decays polynomially with increasing system size, i.e., even in the presence of interactions integrability fundamentally changes the nature of the fluctuations of the eigenstate entanglement entropy. Previous studies had found qualitatively similar enhanced fluctuations of the entanglement entropy in quadratic models~\cite{vidmar_hackl_17, hackl_vidmar_19, lydzba_rigol_20, lydzba_rigol_21} and in Gaussian pure states~\cite{bianchi_hackl_21, bianchi_hackl_22}.

In what follows, by entanglement entropy of a pure state we mean the bipartite entanglement entropy. It is calculated dividing a lattice with $L$ sites into a subsystem $A$ with $L_A$ connected sites and its complement $B$ with $L_B = L - L_A$ connected sites. To compute the entanglement entropy of a pure state $\ket{\psi}$ in subsystem $A$, one first traces out the complement $B$ to obtain the mixed state in $A$
\begin{equation}\label{eq:reduced_matrix}
    \hat{\rho}_A=\Tr_B\dyad{\psi},
\end{equation}
and then computes the von Neumann entropy in subsystem $A$
\begin{equation}\label{eq:reduced_matrix2}
    S_A=-\Tr\qty(\hat{\rho}_A\ln\hat{\rho}_A).
\end{equation}
We calculate the average entanglement entropy $\bar S_A$ in mid-spectrum eigenstates of the spin-$\frac{1}{2}$ $XYZ$ chain.

There are two main reference analytical results with which we contrast our numerical results for $\bar S_A$. The first one is Page's result in the limit of large system sizes
\begin{equation}\label{eq:page_leading}
\langle S_A \rangle = L_A\ln2-\frac{1}{2}\delta_{f,\frac{1}{2}} + o(1),\quad \text{for} \quad f\leq \frac{1}{2},
\end{equation}
where $f=L_A/L$ and $o(1)$ means terms that vanish in the thermodynamic limit. The result for $f>1/2$ follows after changing $L_A\rightarrow L-L_A$ in Eq.~\eqref{eq:page_leading}. We refer to $f$ as the subsystem fraction. As $L \rightarrow \infty$, Page's result differs from the maximal entropy $L_A\ln2$ only at $f=1/2$, and the correction is $O(1)$. 

The other analytic reference is the equivalent of Page's result for Gaussian states, which for $f\leq \frac{1}{2}$ reads~\cite{bianchi_hackl_21, bianchi_hackl_22}
\begin{eqnarray}\label{eq:entropy_quadratic}
    \langle S_A \rangle_G&=& L_A \left[\ln{2} - 1 + (1-f^{-1})\ln(1-f)\right] \nonumber \\
		&& +\frac{1}{2}f+\frac{1}{4}\ln{(1-f)}\,+\,o(1)\,,
\end{eqnarray}
where the leading volume-law term was first found in Ref.~\cite{lydzba_rigol_20} in the context of quantum-chaotic quadratic models. The result for $f>\frac{1}{2}$ follows after changing $L_A\rightarrow L-L_A$ and $f \rightarrow 1-f $ in Eq.~\eqref{eq:entropy_quadratic}. Note that, in contrast with Eq.~\eqref{eq:page_leading}, the coefficient of the volume in Eq.~\eqref{eq:entropy_quadratic} not only is not maximal but actually depends on the subsystem fraction $f$. At $f=1/2$ one finds that as $L\rightarrow\infty$~\cite{lydzba_rigol_20, liu_chen_18}
\begin{equation}\label{eq:entropy_quadratic:1/2}
    \frac{\langle S_A \rangle_G}{\frac{L}{2}\ln{2}} = 2 - (\ln 2)^{-1} \approx 0.5573.
\end{equation}
In the opposite limit, $f \rightarrow 0$, $\langle S_A \rangle_G \simeq L_A \ln{2}$ so it matches Page's result.

We also contrast our results against those for the average eigenstate entanglement entropy of translationally invariant noninteracting fermions in one dimension, $\langle S_A \rangle_T$. No close expression for $\langle S_A \rangle_T$ is known, but tight bounds were obtained in Refs.~\cite{vidmar_hackl_17, hackl_vidmar_19}. $\langle S_A \rangle_T$ was computed numerically in Ref.~\cite{vidmar_hackl_17}, and extrapolating the results for finite systems at $f=1/2$ resulted in the following estimate in the thermodynamic limit~\cite{bianchi_hackl_22}
\begin{equation}\label{eq:entropy_noninteracting}
    \frac{\langle S_A \rangle_T}{\frac{L}{2}\ln{2}} = 0.5378(1),
\end{equation}
which is slightly smaller than the result for Gaussian states in Eq.~\eqref{eq:entropy_quadratic:1/2}. This points towards a lack of universality of the leading volume-law term in quadratic systems that is yet to be understood.

The presentation is organized as follows. In Sec.~\ref{sec:models}, we introduce the spin-$\frac{1}{2}$ $XYZ$ chain Hamiltonian and discuss some of its key symmetries. Subsequently, in Sec.~\ref{sec:param_space}, we explore the parameter space of this model to find an optimal set of points to carry out the finite-size scaling analyses of the average entanglement entropy. In Sec.~\ref{sec:finite_size}, we report our results for the average and the normalized standard deviation of the eigenstate entanglement entropy at $XYZ$ points without special symmetries, and at $XXZ$ and SUSY points. A summary of our results is presented in Sec.~\ref{sec:summary}.

\section{Model and Symmetries}\label{sec:models}

The Hamiltonian of the integrable spin-$\frac{1}{2}$ $XYZ$ chain with periodic boundary conditions reads
\begin{subequations}\label{eq:model:xyz}
    \begin{align}
        &\hat H_\text{XYZ}=\sum_{\ell=1}^LJ_x\hat{s}_\ell^x\hat{s}_{\ell+1}^x+J_y\hat{s}_\ell^y\hat{s}_{\ell+1}^y+J_z\hat{s}_\ell^z\hat{s}_{\ell+1}^z,\\
    &J_x=1-\eta,\qquad J_y=1+\eta,\qquad J_z=\Delta,
    \end{align}
\end{subequations}
where $\hat{s}_\ell^\alpha \doteq \frac{\hbar}{2} \sigma^\alpha_\ell$, with $\alpha=x,y,z$ and $\{\sigma^\alpha\}$ being the Pauli matrices, is the $\alpha$ spin-$\frac{1}{2}$ operator at site $\ell$. The spin-$\frac{1}{2}$ $XYZ$ chain has been thoroughly studied in the literature since the seminal works by Baxter~\cite{Baxter1971, Baxter1972a, Baxter1972b, Baxter1973a, Baxter1973b, Baxter1973c}, who introduced the analytical tools to solve it. 

When $\eta=0$, the Hamiltonian above is that of the spin-$\frac{1}{2}$ $XXZ$ chain, which exhibits U(1) symmetry,
\begin{equation}\label{eq:model:xxz}
    \hat H_\text{XXZ}=\sum_{\ell=1}^L\hat{s}_\ell^x\hat{s}_{\ell+1}^x+\hat{s}_\ell^y\hat{s}_{\ell+1}^y+\Delta\hat{s}_\ell^z\hat{s}_{\ell+1}^z,
\end{equation}
and whose solution dates back to the seminal work of Bethe~\cite{Bethe_31} for the isotropic ($\Delta=1$) case.

Both models, to which we refer to in what follows as the $XYZ$ and $XXZ$ models, have numerous point symmetries that we resolve in order to carry out our full exact diagonalization study. First, because of translational invariance, we block-diagonalize the Hamiltonian using total quasimomentum $k$ eigenkets. For $k=0,\pi$, we further resolve the reflection symmetry present. 

To diagonalize the $XYZ$ model, we take advantage of the fact that it commutes with the parity operators $\mathcal{P}^\alpha=\prod_\ell\sigma^\alpha_\ell$ along the three axes. When the lattice has an even number of sites, the parity operators $\mathcal{P}^\alpha$ for different axes commute with each other. Consequently, the Hamiltonian eigenstates are simultaneous eigenstates of $\mathcal{P}^{x}$, $\mathcal{P}^{y}$, and $\mathcal{P}^{z}$. Since each parity operator can be written as a product of the other two, because any Pauli matrix can be written as a product of the other two, we only need to resolve two of the parities in our numerical calculations. This means that, for $k\neq 0,\pi$, the number of states in each block that needs to be diagonalized is $\approx 2^L/(4L)$, while for $k = 0,\pi$ it is $\approx 2^L/(8L)$. When the lattice has an odd number of sites, the parity operators $\mathcal{P}^\alpha$ for different axes do not commute with each other. Hence, the Hamiltonian eigenstates are only simultaneous eigenstates of one of the parity operators. Consequently, the Hilbert space dimensions of the matrices that need to be diagonalized are $\approx 2^L/(2L)$ for $k\neq 0,\pi$ and $\approx 2^L/(4L)$ for $k= 0,\pi$. After implementing all those point symmetries, we are able to fully diagonalize chains with up to $L=22$ sites. 

In addition to the aforementioned parity symmetries, the $XXZ$ model exhibits a U(1) symmetry. This means that it also conserves the total magnetization, $\hat S^z=\sum_{\ell}\hat s_\ell^z$. We split the total Hilbert space of this model in sectors with fixed magnetization $S^z$, whose dimensions are 
\[
D(S^z)=\binom{L}{S^z+L/2}.
\]
In such sectors $\mathcal{P}^z$ acts trivially, and the only remaining symmetry is the parity symmetry associated with $\mathcal{P}^{x}$ (or $\mathcal{P}^{y}$) at zero magnetization ($S^z=0$). This means that for $S^z\neq 0$ the number of states in each block diagonalized is $\sim D(S^z)/L$ for $k\neq 0,\pi$ and $\sim D(S^z)/(2L)$ for $k= 0,\pi$. At zero magnetization, thanks to the $\mathcal{P}^{x}$ symmetry, the number of states in each block diagonalized is $\sim D(0)/(2L)$ for $k\neq 0,\pi$ and $\sim D(0)/(4L)$ for $k= 0,\pi$. For the $XXZ$ model we are able to fully diagonalize chains with up to $L=24$ sites.

Finally, both the $XYZ$ and $XXZ$ models exhibit SUSY when~\cite{fendley1992, Fendley2010, Fendley2012}
\begin{equation}\label{eq:xyz:susy2}
    J_xJ_y+J_xJ_z+J_yJ_z=0,
\end{equation}
or, using our choice of parameters in Eq.~(\ref{eq:model:xyz}), when
\begin{equation}\label{eq:xyz:susy1}
    \Delta=-\frac{1-\eta^2}{2}.
\end{equation}
More specifically, in Ref.~\cite{Fendley2012} it was shown that the $XYZ$ model exhibits $\mathcal{N}=(2,2)$ lattice SUSY. For chains with an even number of sites, one can extend the condition in Eq.~(\ref{eq:xyz:susy1}) by carrying out a spin-parity transformation on even sites only~\cite{Paritynote}, leading to
\begin{equation}\label{eq:xyz:susy}
    \Delta=\pm\frac{1-\eta^2}{2}\quad\vee\quad\Delta=\pm\frac{1-\eta^2}{2\eta}.
\end{equation}
The SUSY condition in Eq.~\eqref{eq:xyz:susy1} is a particular instance of the root-of-unity symmetry~\cite{Hagendorf2013,Hagendorf2018}.

For our study of the standard deviations of the eigenstate entanglement entropy, we also consider quantum-chaotic interacting Hamiltonians
\begin{eqnarray}\label{eq:model:xyzQC}
    \hat H^\text{QC}_\text{XYZ} &=& \hat H_\text{XYZ}+\hat H^\text{nn}_\text{XYZ},\\
    \hat H^\text{QC}_\text{XXZ} &=& \hat H_\text{XXZ}+\hat H^\text{nn}_\text{XXZ}, \label{eq:model:xxzQC}
\end{eqnarray}
where
\begin{eqnarray*}\label{eq:model:xyznn}
    \hat H^\text{nn}_\text{XYZ}\!\!&=&\!\!J_2\sum_{\ell=1}^L(1-\eta)\hat{s}_\ell^x\hat{s}_{\ell+2}^x+(1+\eta)\hat{s}_\ell^y\hat{s}_{\ell+2}^y+\Delta \hat{s}_\ell^z\hat{s}_{\ell+2}^z,\\
    \hat H^\text{nn}_\text{XXZ}\!\!&=&\!\!J_2\sum_{\ell=1}^L\hat{s}_\ell^x\hat{s}_{\ell+2}^x+\hat{s}_\ell^y\hat{s}_{\ell+2}^y+\Delta\hat{s}_\ell^z\hat{s}_{\ell+2}^z.\label{eq:model:xxznn}
\end{eqnarray*}
In those calculations, we take $\eta$ and $\Delta$ to be the same in $\hat H_\text{XYZ}$ and $\hat H^\text{nn}_\text{XYZ}$, as well as $\Delta$ to be the same in $\hat H_\text{XXZ}$ and $\hat H^\text{nn}_\text{XXZ}$. Both, for $\hat H^\text{nn}_\text{XYZ}$ and $\hat H^\text{nn}_\text{XXZ}$, we take $J_2=2$ as in Ref.~\cite{Kliczkowski_Swietek2023}.

\section{Parameter space}\label{sec:param_space}

In addition to the U(1) and SUSY lines, the integrable $XYZ$ model exhibits other special points and lines that are of no particular interest to us here, and which can affect the outcome of the numerical calculations if we choose parameters that are too close to them. For $\Delta=0$ and $\eta=\pm 1$ we have the (classical) Ising model. For $\Delta=0$ and $\eta \neq \pm 1$, or $\Delta\neq 0$ and $\eta = \pm 1$, we have the so-called $XY$ model, which is mappable onto a quadratic model (free fermions for $\Delta=\eta=0$). For $\Delta=\pm1$ and $\eta=0$ we have the isotropic Heisenberg model, which has $SU(2)$ symmetry and whose eigenstate entanglement entropy was recently studied in Ref.~\cite{patil2023}.

Since the $XYZ$ is a local interacting model, we expect the density of states to be Gaussian for very large systems, as it is in local quantum-chaotic interacting models~\cite{brody_flores_81}. To quantify how Gaussian the density of states is in the finite chains that we study, we compute the ratio
\begin{equation}\label{eq:gaussianity}
    \Gamma_E=\frac{\overline{\abs{E_\alpha}^2}}{\overline{\abs{E_\alpha}}^2},
\end{equation}
where $E_\alpha$ are the eigenenergies of the Hamiltonian. For a Gaussian distribution, $\Gamma^G_E=\frac{\pi}{2}$. 

\begin{figure}[!t]
\includegraphics[width=0.98\linewidth]{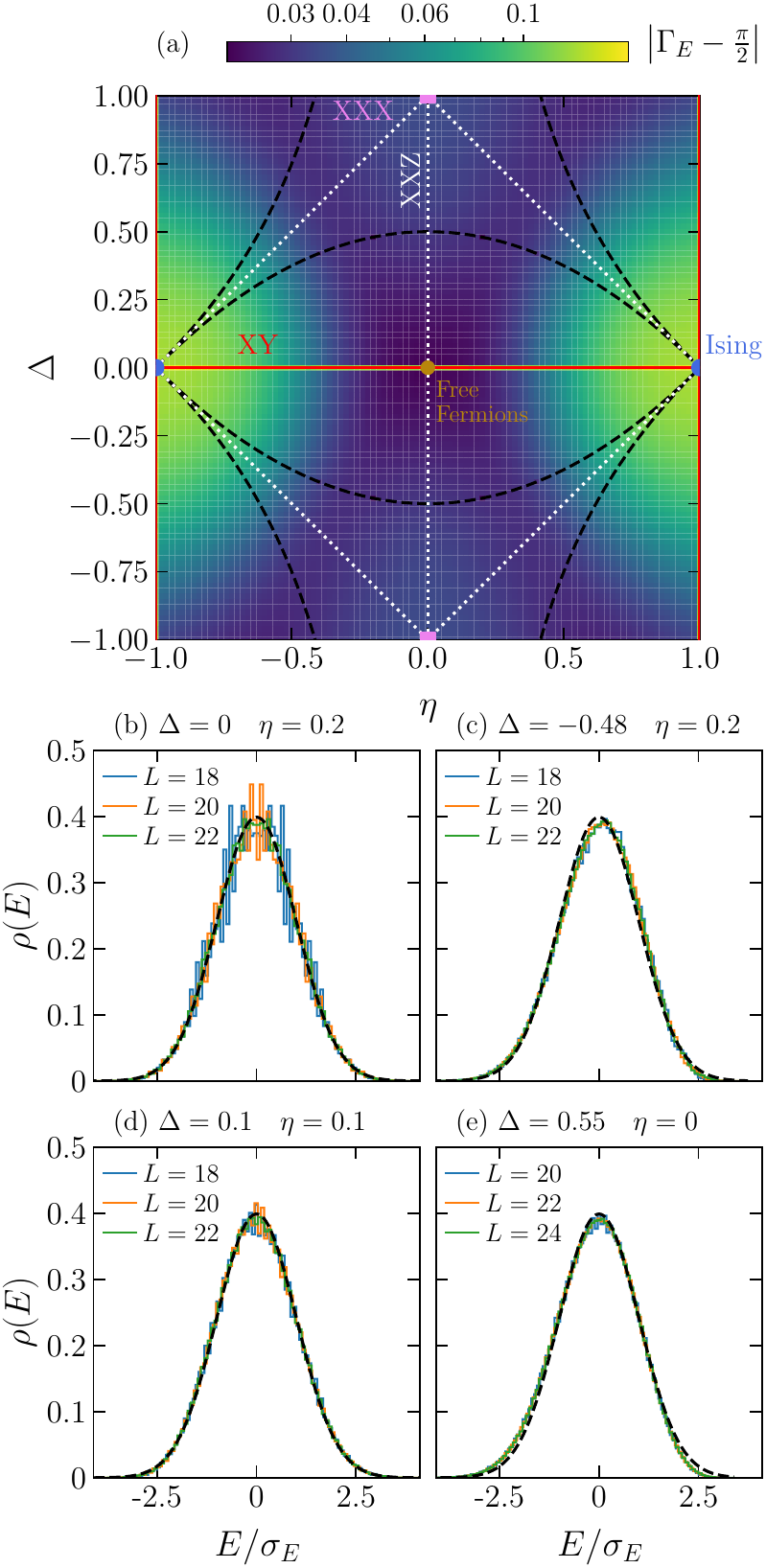}
\vspace{-0.15cm}
\caption{(a) $|\Gamma_E-\frac{\pi}{2}|$ as a function of $\Delta$ and $\eta$ in the $XYZ$ model for $L=18$. Dots are used to label the Ising, free fermion, and Heisenberg ($XXX$) points, while the $XY$, U(1)~\cite{U1note}, and SUSY lines are marked using continuous (red), dotted (white), and dashed (black) lines, respectively. The lower panels show the normalized density of states at (b) $XY$, (c) SUSY, (d) $XYZ$, and (e) $XXZ$ in the $S_z=0$ sector points plotted as functions of rescaled energy $E/\sigma_E$, where $\sigma_E^2=\overline{E_\alpha^2}-\overline{E_\alpha}^2$ is the variance of the energy. As a guide to the eye, we plot Gaussian distributions with the same $\sigma_E^2$ (dashed lines).}
\label{fig:gaussianity}
\end{figure}

In Fig.~\ref{fig:gaussianity}(a) we show the difference $|\Gamma_E-\frac{\pi}{2}|$ across the explored parameter space for the $XYZ$ model. Small values of $|\Gamma_E-\frac{\pi}{2}|$ indicate closer to Gaussian distributions and weaker finite-size effects. All the special lines and points with the symmetries discussed before are marked in the diagram. We find the largest deviations from a Gaussian density of states, accompanied by the strongest finite-size effects, close to the Ising points. As the system moves away from those points the density of states approaches a Gaussian distribution. We then find slightly larger deviations from a Gaussian density of states close to the $XY$ and SUSY lines, see Figs.~\ref{fig:gaussianity}(b) and~\ref{fig:gaussianity}(c), respectively, when compared to a typical $XYZ$ or $XXZ$ point, see Figs.~\ref{fig:gaussianity}(d) and~\ref{fig:gaussianity}(e). The parameters for which density of states are shown in Figs.~\ref{fig:gaussianity}(c)--\ref{fig:gaussianity}(e) are used later for our finite-size scaling analyses of the eigenstate entanglement entropy, specifically, those in Fig.~\ref{fig:gaussianity}(c) are used in Sec.~\ref{sec:susy}, those in Fig.~\ref{fig:gaussianity}(d) are used in Sec.~\ref{sec:scaling:xyz}, and those in Fig.~\ref{fig:gaussianity}(e) are used in Sec.~\ref{sec:scaling:xxz}.

\begin{figure}[!t]
\includegraphics[width=0.93\columnwidth]{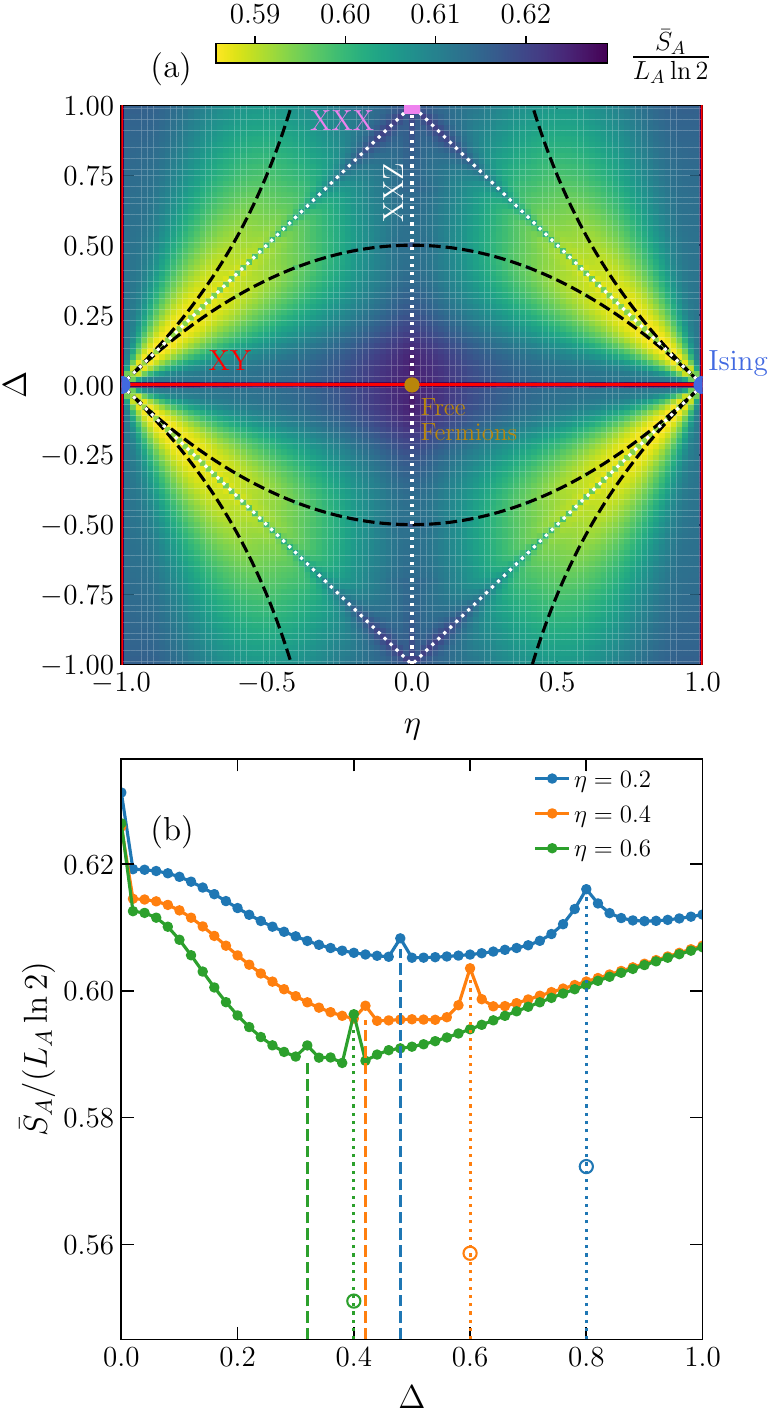}
\vspace{-0.2cm}
\caption{(a) Average eigenstate entanglement entropy $\bar{S}_A$, calculated over all the eigenstates of the $XYZ$ model, as a function of $\Delta$ and $\eta$ for $L=18$ at $L_A=L/2$. Dots are used to label the Ising, free fermion, and Heisenberg ($XXX$) points, while the $XY$, U(1), and SUSY lines are marked using continuous (red), dotted (while), and dashed (black) lines, respectively. (b) Cuts in (a) along specific values of $\eta$ for $\Delta\geq0$. The vertical dashed lines indicate SUSY points, while the dotted lines indicate U(1) points. At the U(1) points, the empty symbols show $\bar{S}_A$ computed after breaking the Hilbert space into sectors with fixed total magnetization.}
\label{fig:entropy_map}
\end{figure}

In Fig.~\ref{fig:entropy_map}(a) we show results for the average eigenstate entanglement entropy $\bar{S}_A$ at $f=1/2$ across the explored parameter space for the $XYZ$ model. The results were obtained averaging the eigenstate entanglement entropy over all eigenstates of the $XYZ$ model, after resolving the space and the two spin-parity symmetries. Figure~\ref{fig:entropy_map}(b) shows the results of cuts in Fig.~\ref{fig:entropy_map}(a) along specific values of $\eta$. For all parameters considered, $\bar{S}_A$ is significantly lower than the $L_A \ln 2$ result expected for quantum-chaotic interacting models. In Fig.~\ref{fig:entropy_map}(a) one can see that the lowest values of $\bar{S}_A$ occur between the SUSY lines as one approaches the Ising points, while the largest values of $\bar{S}_A$ occur about the $\Delta=0$ line. Figure~\ref{fig:entropy_map}(b) further reveals sharp features in $\bar{S}_A$ at the U(1) and SUSY lines, which are the result of degeneracies present in the energy spectrum. This can be understood as the exact diagonalization returns a superposition of the Hamiltonian eigenstates within the degenerate subspace. Resolving the U(1) symmetry, see the empty points in Fig.~\ref{fig:entropy_map}(b), reduces $\bar{S}_A$ significantly in the small system sizes that we study. In Sec.~\ref{sec:susy} we discuss how degeneracies affect the results for $\bar{S}_A$ in the presence of SUSY.

\section{Average entanglement entropy\label{sec:finite_size}}

Here we study the scaling $\bar S_A$ and $\Sigma_S$ of midspectrum energy eigenstates in the $XYZ$ model. Our focus is on the subsystem fraction $f=L_A/L=1/2$. To reduce finite-size effects (energy eigenstates at the edges of the energy spectrum are responsible for the largest finite-size corrections), the averages $\bar{S}_A$ and normalized standard deviations $\Sigma_S$ are computed over fractions $\nu\leq 0.5$ of midspectrum energy eigenstates, i.e., we compute $\bar{S}_A(\nu)$ and $\Sigma_S(\nu)$. The leading term of $\bar{S}_A(\nu)$, our interest here, is independent of $\nu$~\cite{Kliczkowski_Swietek2023}.

Following Ref.~\cite{Kliczkowski_Swietek2023}, we calculate the mean energy $\bar{\rm E}$ within each symmetry sector, which in our finite and small systems is different from the global mean energy $\bar E$ calculated over the entire Hilbert space. To compute the averages, we use a fraction $\nu$ of midspectrum eigenstates about $\bar{\rm E}$. The results reported are the weighted averages over all symmetry sectors, with the weight determined by the Hilbert space dimension of each sector relative to the total Hilbert space.

\begin{figure}[!t]
\includegraphics[width=0.98\columnwidth]{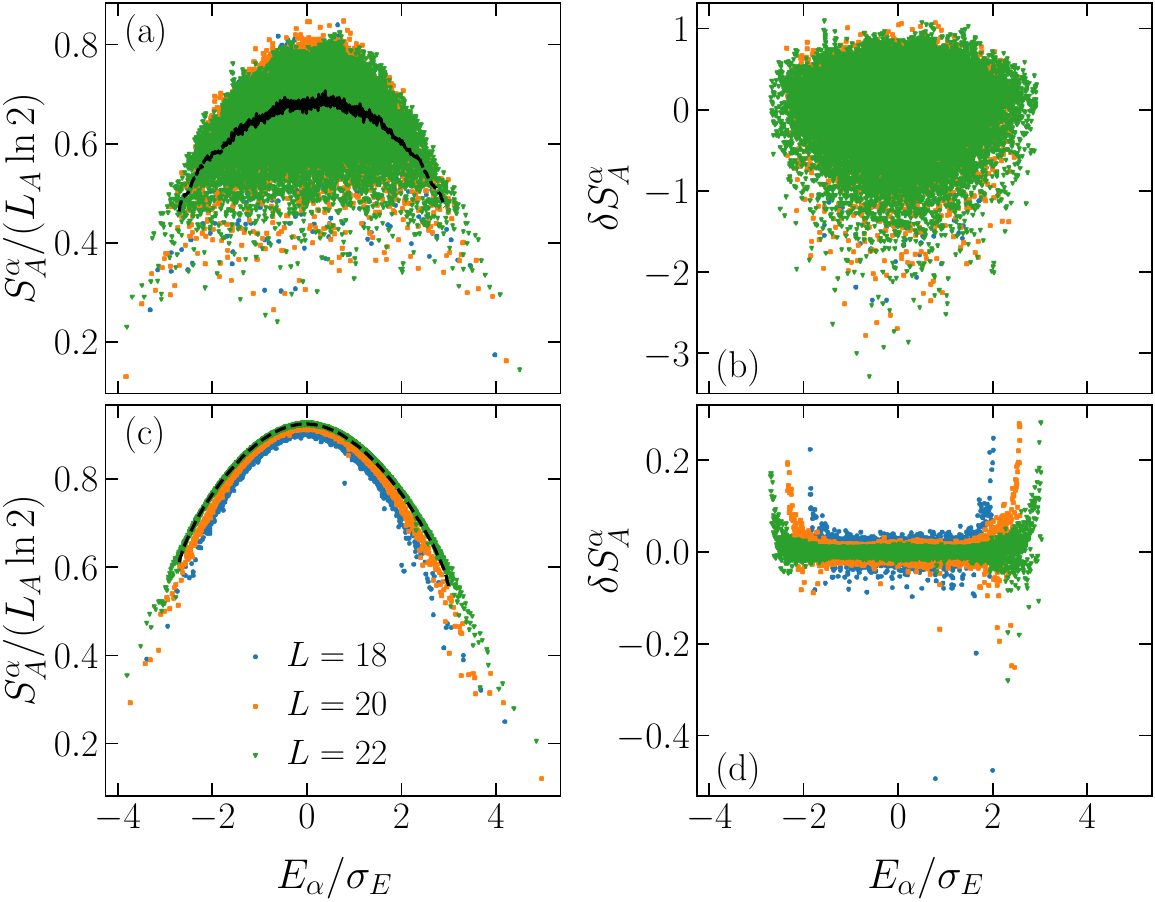}
\vspace{-0.15cm}
\caption{Eigenstate entanglement entropy as a function of the eigenenergies for the $XYZ$ model with $\Delta=\eta=0.2$. [(a) and (b)] integrable interacting point, see Eq.~\eqref{eq:model:xyz}, and [(c) and (d)] quantum-chaotic interacting point (with $J_2=2$), see Eq.~\eqref{eq:model:xyzQC}. Results are shown for three system sizes in the sector with $k=0$, and eigenvalue 1 for the reflection and parity in $x$ and $z$ symmetries. Panels (b) and (d) show the differences $\delta S_\alpha$ between the eigenstate entanglement entropy and the running average [see Eq.~\eqref{app:eq:structurless_entropy}]. The running average [with $\kappa=100$, see Eq.~\eqref{app:eq:moving_average}] for the largest system size is shown in (a) and (c) as a dashed black line.}
\label{fig:entropy_structure}
\end{figure}

We also study the scaling of the standard deviation of the eigenstate entanglement entropy. For such scalings to be meaningful, one needs to account for the fact that, for local Hamiltonians such as those in Eqs.~\eqref{eq:model:xyz},~\eqref{eq:model:xxz},~\eqref{eq:model:xyzQC},~and~\eqref{eq:model:xxzQC}, the eigenstate entanglement entropy displays a structure as a function of eigenenergies [see Figs.~\ref{fig:entropy_structure}(a) and~\ref{fig:entropy_structure}(c)]. The importance of removing the structure when analyzing fluctuations was already recognized in studies of the matrix elements of observables in the context of quantum chaos and the eigenstate thermalization hypothesis~\cite{beugeling_moessner_14, mondaini_rigol_17, jansen_stolpp_19, mierzejewski_vidmar_20}. In Ref.~\cite{Kliczkowski_Swietek2023}, we used such a structure in quantum-chaotic interacting models to derive an expression that describes the dependence of the $O(1)$ term in $\bar{S}_A(\nu)$ on $\nu$. 

To remove the effect of the energy dependence of the eigenstate entanglement entropy on the standard deviation, we first compute the running average
\begin{equation}\label{app:eq:moving_average}
    \bar{S}_A^\alpha=\frac{1}{\kappa+1}\sum_{\beta=\alpha-\frac{\kappa}{2}}^{\alpha+\frac{\kappa}{2}}S_A^\beta,
\end{equation}
for each eigenstate $\alpha$, see Figs.~\ref{fig:entropy_structure}(a) and~\ref{fig:entropy_structure}(c), and then the deviations 
\begin{equation}\label{app:eq:structurless_entropy}
    \delta S_{A}^\alpha = S_A^\alpha - \bar{S}_A^\alpha,
\end{equation}
see Figs.~\ref{fig:entropy_structure}(b) and~\ref{fig:entropy_structure}(d). For our finite-size scaling analyses, we take $\kappa=20$ for $L<18$ and $\kappa=100$ otherwise.

Our normalized standard deviation is then defined as
\begin{equation}\label{eq:entropy_var}
    \Sigma_{S}(\nu)=\frac{\sqrt{\,\overline{(\delta S_{A}^\alpha)^2}(\nu)}}{L_A\ln2},
\end{equation}
where the average $\overline{(\delta S_{A}^\alpha)^2}(\nu)$ is also computed over a fraction $\nu$ of midspectrum eigenstates. The normalization in $\Sigma_S$ accounts for the fact that $\bar S_A$ is proportional to $L_A$ in the thermodynamic limit.

In Appendix~\ref{app:entropy_structure}, we show that not accounting for the structure of the eigenstate entanglement entropy fundamentally changes the scaling of the normalized standard deviation in quantum-chaotic interacting Hamiltonians.

\subsection{XYZ model\label{sec:scaling:xyz}}

\begin{figure}[!t]
\includegraphics[width=0.98\columnwidth]{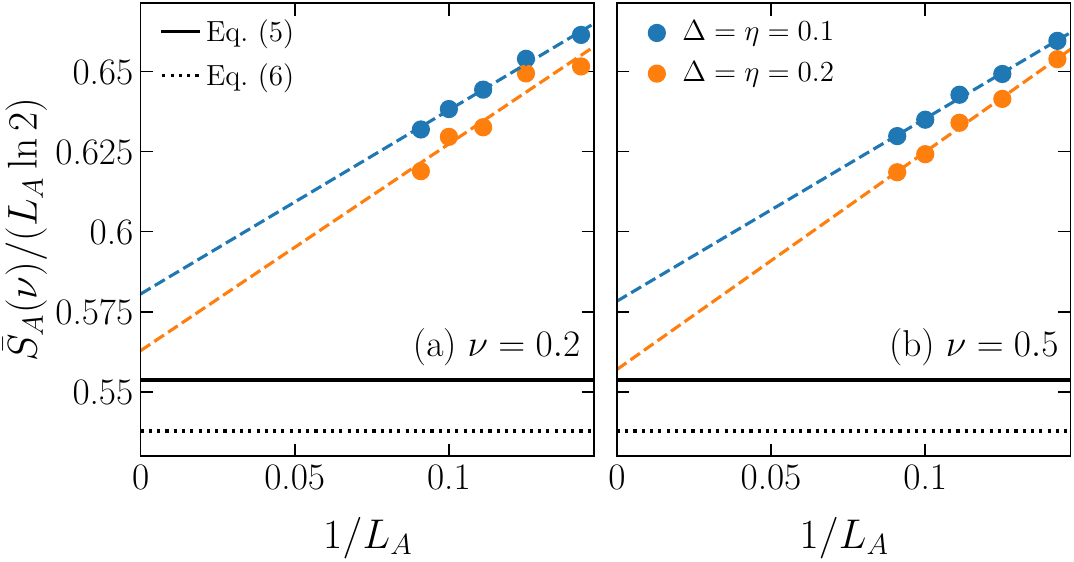}
\vspace{-0.15cm}
\caption{Finite-size scaling analyses of $\bar{S}_A(\nu)$ in the $XYZ$ model for (a) $\nu = 0.2$ and (b) $\nu = 0.5$ at $f=L_A/L=1/2$. We show results for $\Delta = \eta = 0.1$ and $\Delta = \eta = 0.2$. The horizontal lines display $\langle S_A \rangle_G$ [see Eq.~\eqref{eq:entropy_quadratic:1/2}] for the Haar-random average over Gaussian states (solid lines) and $\langle S_A \rangle_T$ [see Eq.~\eqref{eq:entropy_noninteracting}] for the average over translationally invariant noninteracting fermions (dotted lines). The dashed lines show fits of  $a+b/L_A$ to the data.}
\label{fig:xyz_size_scaling}
\end{figure}

For the $XYZ$ model, following our discussion in Sec.~\ref{sec:param_space}, we take the Hamiltonian parameters to be $\Delta = \eta = 0.1$ and $\Delta = \eta = 0.2$. In Fig.~\ref{fig:xyz_size_scaling}, we show results for $\bar{S}_A(\nu)$ for $\nu = 0.2$ [Fig.~\ref{fig:xyz_size_scaling}(a)] and $\nu = 0.5$ [Fig.~\ref{fig:xyz_size_scaling}(b)]. The horizontal solid and dotted lines show $\langle S_A \rangle_G$ for the Haar random average over Gaussian states [see Eq.~\eqref{eq:entropy_quadratic:1/2}] and $\langle S_A \rangle_T$ for the average over translationally invariant noninteracting fermions in one dimension [see Eq.~\eqref{eq:entropy_noninteracting}], respectively. The dashed lines show fits of $a+b/L_A$ to the data.

Our results in Figs.~\ref{fig:xyz_size_scaling}(a) and~\ref{fig:xyz_size_scaling}(b) make apparent that, in contrast with quantum-chaotic interacting models, $\bar{S}_A(\nu)/(L_A\ln 2)$ at $f=1/2$ decreases with increasing system size and reaches a value that is between 0.55 and 0.6 depending on the Hamiltonian parameters. The results for $\nu = 0.2$ in Fig.~\ref{fig:xyz_size_scaling}(a) are very close to those for $\nu = 0.5$ in Fig.~\ref{fig:xyz_size_scaling}(b), as expected for the leading volume-law term (which should be independent of $\nu$). Our results for $\bar{S}_A/(L_A\ln 2)$ indicate that the leading volume-law term in the integrable interacting $XYZ$ chain is different from the maximal value observed in quantum-chaotic interacting models without U(1) symmetry (particle-number conservation)~\cite{Kliczkowski_Swietek2023}, and may depend on the Hamiltonian parameters. This, together with the fact that $\langle S_A \rangle_G$ and $\langle S_A \rangle_T$ are also different, suggests that the leading volume-law term of $\bar{S}_A$ in integrable models depends on the Hamiltonian considered and, for any given Hamiltonian, it may depend on the parameters selected.

\begin{figure}[!b]
\includegraphics[width=0.98\columnwidth]{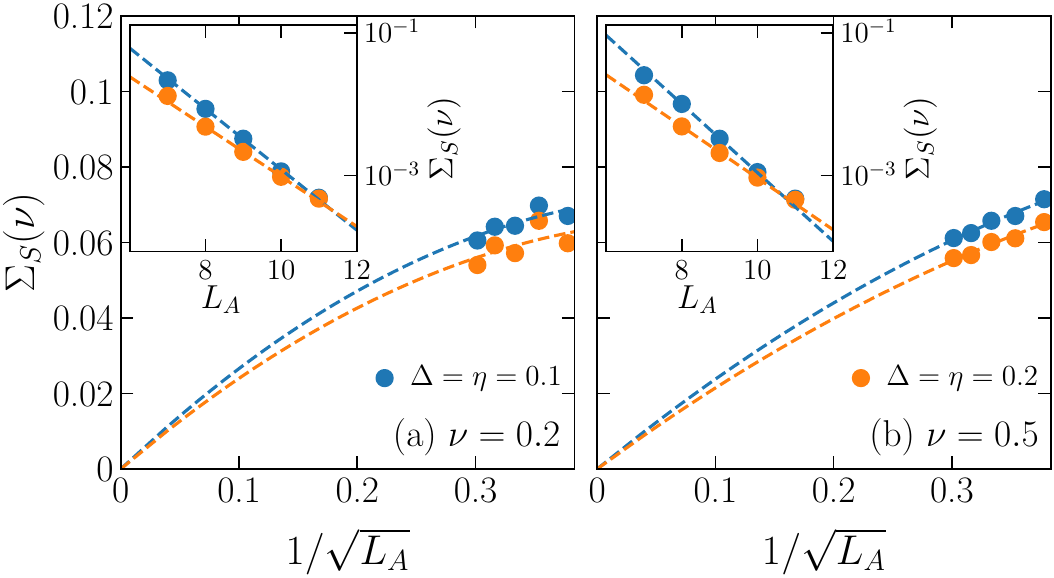}
\vspace{-0.15cm}
\caption{Finite-size scaling analyses of $\Sigma_S(\nu)$ in the $XYZ$ model for (a) $\nu = 0.2$ and (b) $\nu = 0.5$ at $f=L_A/L=1/2$. We show results for $\Delta = \eta = 0.1$ and $\Delta = \eta = 0.2$. The dashed lines display fits of $a/\sqrt{L_A}+b/L_A$ to the data. (Insets) Corresponding results at quantum-chaotic interacting points with the same values of $\Delta$ and $\eta$, and $J_2=2$, see Eq.~\eqref{eq:model:xyzQC}. The dashed lines display fits of $a\exp(-bL_A)$ to the data for the four largest system sizes.}
\label{fig:xyz_size_scalingvariance}
\end{figure}

In the main panels in Fig.~\ref{fig:xyz_size_scalingvariance}, we show results for $\Sigma_S(\nu)$ for $\nu = 0.2$ [Fig.~\ref{fig:xyz_size_scalingvariance}(a)] and $\nu = 0.5$ [Fig.~\ref{fig:xyz_size_scalingvariance}(b)]. The dashed lines display fits of $a/\sqrt{L_A}+b/L_A$ to the data. In the insets, we show results for the corresponding to quantum-chaotic interacting points as per Eq.~\eqref{eq:model:xyzQC}. The dashed lines in the insets display fits of $a\exp(-bL_A)$ to the data. The contrast between the scalings seen in the main panels and the insets in Figs.~\ref{fig:xyz_size_scalingvariance}(a) and~\ref{fig:xyz_size_scalingvariance}(b) unveil another fundamental difference between the entanglement properties of eigenstates of integrable interacting models and of quantum-chaotic interacting models. In the former $\Sigma_S$ vanishes polynomially with increasing system size, while in the latter it vanishes exponentially. Our results indicate that for large systems sizes, at $f=1/2$, $\Sigma_S\propto 1/\sqrt{L_A}$ in the $XYZ$ chain as found in Ref.~\cite{vidmar_hackl_17} for translationally invariant free fermions in one dimension. Such a decay is slower than the $\Sigma_S\propto 1/L_A$ found in Ref.~\cite{lydzba_rigol_20} for random quadratic Hamiltonians as well as in Refs.~\cite{bianchi_hackl_21, bianchi_hackl_22} for the Haar-random average over Gaussian states. The fact that $\Sigma_S$ vanishes in the thermodynamic limit, even if polynomially in $L_A$, means that $\bar{S}_A$ is also the typical value of the entanglement entropy in eigenstates of integrable interacting Hamiltonians.

\subsection{$XXZ$ model\label{sec:scaling:xxz}}

\begin{figure}[!b]
\includegraphics[width=0.98\columnwidth]{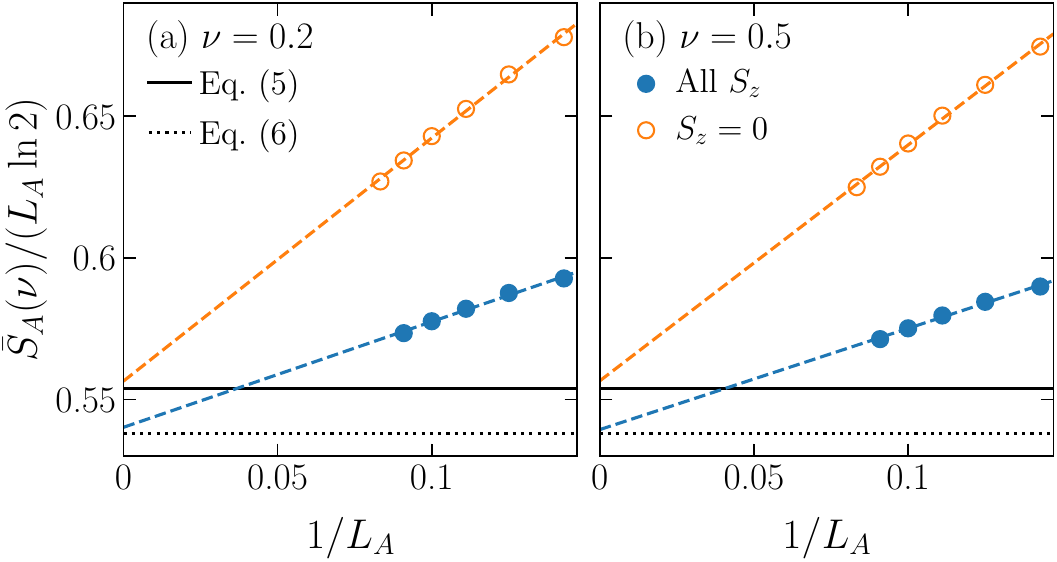}
\vspace{-0.15cm}
\caption{Same as Fig.~\ref{fig:xyz_size_scaling} for the $XXZ$ model with $\Delta =0.55$. Results are shown for the average over the zero magnetization sector (empty symbols) as well as over all magnetization sectors (filled symbols).}
\label{fig:xxz_size_scaling}
\end{figure}

In Fig.~\ref{fig:xxz_size_scaling} we show scaling analyses of $\bar{S}_A(\nu)$ for the $XXZ$ model (with $\Delta =0.55$) at $f=L_A/L=1/2$, which parallel those reported in Fig.~\ref{fig:xyz_size_scaling} for the $XYZ$ model. In each panel of Fig.~\ref{fig:xxz_size_scaling}, we report the averages obtained over all magnetization sectors (filled symbols) and the averages in the sector with zero magnetization $S^z = 0$ (empty symbols). 

As for the $XYZ$ model, the results in Figs.~\ref{fig:xxz_size_scaling}(a) and~\ref{fig:xxz_size_scaling}(b) for $\bar{S}_A(\nu)$ when $\nu=0.2$ and $\nu=0.5$, respectively, are very close to each other. At fixed $\nu$, on the other hand, $\bar{S}_A(\nu)$ in the zero magnetization sector is larger than when the average is carried out over all magnetization sectors. The fits of $a+b/L_A$ to the data in Figs.~\ref{fig:xxz_size_scaling}(a) and~\ref{fig:xxz_size_scaling}(c) make apparent that the differences between those two averages decrease with increasing system size. In the thermodynamic limit, since the overwhelming majority of the Hilbert space has average zero magnetization per site, one expects $\bar{S}_A$ to exhibit the same volume-law term for both averages, with the differences emerging in the subleading corrections, see, e.g., the pedagogical discussions in Ref.~\cite{bianchi_hackl_22}. Hence, the differences between our fitted parameter $a$ for both averages (the crossing points in the $y$ axes) are expected to be the result of subleading corrections not accounted for in the fits to the small system sizes available to us. Those differences give an idea of our error in the estimation of the coefficient of the volume-law term in the $XXZ$ model. Our results in Figs.~\ref{fig:xxz_size_scaling}(a) and~\ref{fig:xxz_size_scaling}(b) complement those in Ref.~\cite{leblond_mallayya_19}, where $\bar{S}_A$ was reported only in the zero magnetization sector and only for (the more affected by finite-size effects) fraction $\nu=1$. 

\begin{figure}[!t]
\includegraphics[width=0.98\columnwidth]{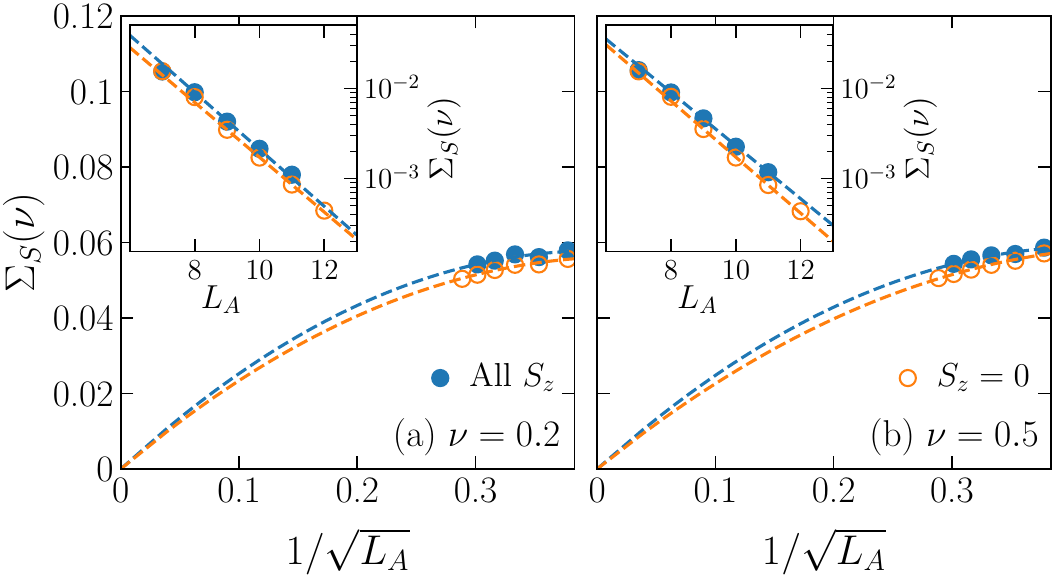}
\vspace{-0.15cm}
\caption{Same as Fig.~\ref{fig:xyz_size_scalingvariance} for the $XXZ$ model with $\Delta =0.55$. Results are shown for the average over the zero magnetization sector (empty symbols) as well as over all magnetization sectors (filled symbols).}
\label{fig:xxz_size_scalingvariance}
\end{figure}

The normalized standard deviations $\Sigma_S(\nu)$ for the $XXZ$ model in Fig.~\ref{fig:xxz_size_scalingvariance} are slightly smaller in the zero magnetization sector than in the average over all magnetization sectors. This is expected given that the zero magnetization sector is the largest sector in the Hilbert space. The results in the main panels in Fig.~\ref{fig:xxz_size_scalingvariance} indicate that for eigenstates of the $XXZ$ chain at $f=1/2$, $\Sigma_S\propto 1/\sqrt{L_A}$ for both averages as $L_A\rightarrow\infty$. For quantum-chaotic energy eigenstates, on the other hand, we find the standard deviations $\Sigma_S$ to decay exponentially in both cases. Hence, $\Sigma_S$ in quantum-chaotic energy eigenstates mirrors the exponential decay found for Haar-random pure states at fixed magnetization for $f=1/2$~\cite{bianchi_hackl_22}.

\subsection{SUSY points\label{sec:susy}}

A remarkable consequence of SUSY in the $XYZ$ and $XXZ$ chains is the formation of degenerate multiplets of eigenstates from different Hilbert spaces. The SUSY charges allow one to map an eigenstate of a chain with $L$ spins onto an eigenstate of a chain with $L\pm1$ spins \cite{Fendley2003, Fendley2012}. In general, SUSY does not imply that there should be degeneracies at any given fixed system size. Degeneracies typically arise due to additional symmetries. We find that, after resolving all the symmetries discussed in Sec.~\ref{sec:models}, there are indeed degeneracies present in the $XYZ$ and $XXZ$ chains at SUSY points, which are a consequence of the roots of unity symmetry. In what follows we study the average entanglement entropy computed over nondegenerate and degenerate energy eigenstates, as well as over only nondegenerate or only degenerate energy eigenstates. To decide whether two or more energy levels are degenerate, since the average level spacing in our calculations is always higher than $10^{-8}$, we round the eigenenergies to 14 significant digits.

\begin{figure}[!t]
\includegraphics[width=0.98\columnwidth]{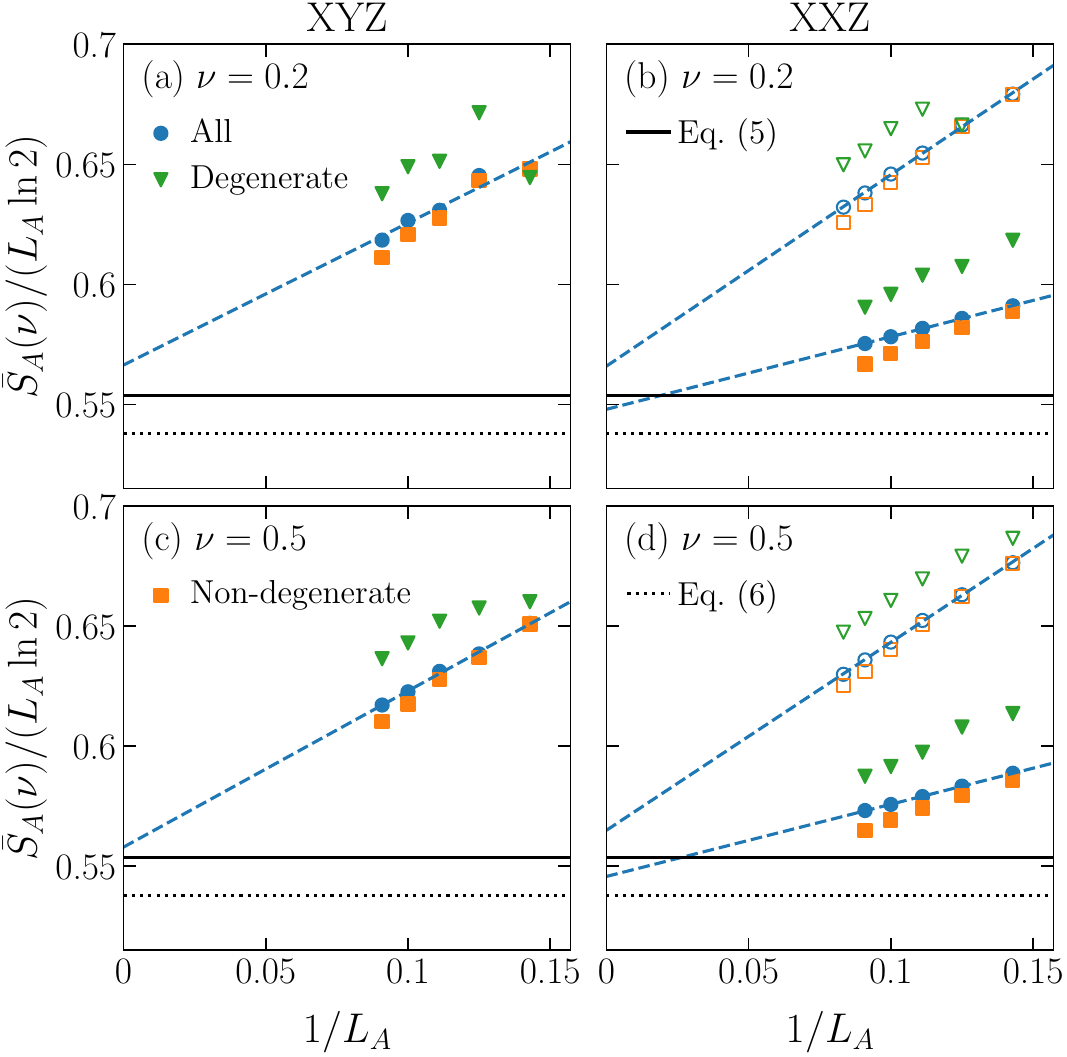}
\vspace{-0.15cm}
\caption{Finite-size scaling analyses of $\bar{S}_A$ at SUSY points for $f=L_A/L=1/2$. Results are reported for the $XYZ$ model with $\eta=0.2$ and $\Delta=-0.48$ for (a) $\nu=0.2$ and (c) $\nu=0.5$, and for the $XXZ$ model with $\Delta=-0.5$ for (b) $\nu=0.2$ and (d) $\nu=0.5$. For the $XXZ$ model, we show results for both the average over the zero magnetization sector (empty symbols) and over all magnetization sectors (filled symbols). For all cases, we show results obtained averaging over degenerate and nondegenerate eigenstates (``All''), as well as averaging over degenerate (``Degenerate'') or nondegenerate (``Nondegenerate'') eigenstates independently. The horizontal lines show $\langle S_A \rangle_G$ from Eq.~\eqref{eq:entropy_quadratic:1/2} and $\langle S_A \rangle_T$ from Eq.~\eqref{eq:entropy_noninteracting}. The dashed lines show fits of $a+b/L_A$ to the data for the average over degenerate and nondegenerate eigenstates (``All'').}
\label{fig:size_scaling_degenerate}
\end{figure}

In Fig.~\ref{fig:size_scaling_degenerate}, we show the scaling of $\bar{S}_A(\nu)$ at SUSY points for the $XYZ$ model [Fig.~\ref{fig:size_scaling_degenerate}(a) for $\nu=0.2$ and Fig.~\ref{fig:size_scaling_degenerate}(c) for $\nu=0.5$] and for the $XXZ$ model [Fig.~\ref{fig:size_scaling_degenerate}(b) for $\nu=0.2$ and Fig.~\ref{fig:size_scaling_degenerate}(d) for $\nu=0.5$], for $f=L_A/L=1/2$. Results for $\bar{S}_A$ are reported for the averages over degenerate and nondegenerate eigenstates (``All''), as well as for the averages over degenerate (``Degenerate'') or nondegenerate (``Nondegenerate'') eigenstates. For the $XXZ$ model, as before, we report the averages obtained over all magnetization sectors (filled symbols) and in the sector with zero magnetization $S^z = 0$ (empty symbols). 

The results obtained for the average over degenerate and nondegenerate eigenstates (``All'') in Figs.~\ref{fig:size_scaling_degenerate}(a),\ref{fig:size_scaling_degenerate}(c) and in Figs.~\ref{fig:size_scaling_degenerate}(b),\ref{fig:size_scaling_degenerate}(d) are similar to those obtained in Figs.~\ref{fig:xyz_size_scaling}(a),\ref{fig:xyz_size_scaling}(b) and in Figs.~\ref{fig:xxz_size_scaling}(a),\ref{fig:xxz_size_scaling}(b), respectively, i.e., SUSY does not appear to introduce any fundamental change in the behavior of the average eigenstate entanglement entropy. A fit of $a+b/L_A$ to the averages over degenerate and nondegenerate eigenstates (``All'') is reported in each panel. One can see that the coefficients $a$ of the leading volume-law term in the absence and presence of U(1) symmetry are close to those obtained in Figs.~\ref{fig:xyz_size_scaling} and~\ref{fig:xxz_size_scaling}, respectively.

The results for the average over only nondegenerate states in all panels in Fig.~\ref{fig:size_scaling_degenerate} is slightly below the corresponding overall average, while those for the average only over degenerate states are above. The latter is expected as, in the degenerate subspaces, the numerical diagonalization returns superpositions of the true Hamiltonian eigenstates. One can see in all panels in Fig.~\ref{fig:size_scaling_degenerate} that the independent averages suffer from stronger finite-size effects than those of the overall averages. Specifically, the averages over degenerate states are nonmonotonic within the system sizes considered, and the averages over nondegenerate eigenstates exhibit a stronger correction beyond $b/L_A$. For this reason, we do not report fits for the independent averages.

The normalized standard deviations $\Sigma_S(\nu)$ at SUSY points, when considering degenerate and nondegenerate eigenstates as well as only nondegenerate eigenstates, resemble the results reported in Figs.~\ref{fig:xyz_size_scalingvariance}(a),~\ref{fig:xyz_size_scalingvariance}(b) and in Figs.~\ref{fig:xxz_size_scalingvariance}(a),~\ref{fig:xxz_size_scalingvariance}(b) for the $XYZ$ and $XXZ$ model, respectively, so we do not report them here. When considering only degenerate eigenstates, as for $\bar{S}_A$ in Fig.~\ref{fig:size_scaling_degenerate}, we find strong finite-size effects that, for the smallest system sizes, result in a nonmonotonic behavior of $\Sigma_S$ with changing system size.

Our results in this section open the question, which will need to be addressed by means beyond full exact diagonalization, of whether the typical degenerate and nondegenerate eigenstates at SUSY points exhibit different entanglement entropies.

\section{Summary}\label{sec:summary}

In this work we report a state-of-the-art numerical study of the average and the standard deviation of the entanglement entropy of energy eigenstates of the integrable spin-$\frac{1}{2}$ $XYZ$ chain. We implement all point symmetries, including the spin parity symmetries of the $XYZ$ chain, to diagonalize systems with up to $L=22$ ($L=24$) sites in the absence (presence) of U(1) symmetry. Our focus was on the subsystem fraction $f=1/2$. For the $XXZ$ model, we report averages over all magnetization sectors and in the sector with zero magnetization $S^z = 0$.

We find that, for all parameters considered, the average entanglement entropy of energy eigenstates of the integrable spin-$\frac{1}{2}$ $XYZ$ chain is below Page's prediction. In all instances we find indications that the leading volume law term is close to the one predicted for the Haar-random average over Gaussian states as well as to the average eigenstate entanglement entropy of translationally invariant noninteracting fermions in one dimension. Our results suggest that the average eigenstate entanglement entropy can be widely used to distinguish integrable and nonintegrable interacting systems. Previous studies had only considered systems with U(1)~\cite{leblond_mallayya_19} or $SU(2)$~\cite{patil2023} symmetry. Our findings are specially important in the context of SUSY points, because the presence of degeneracies could have produced a qualitative change in the behavior of the average entanglement entropy. An interesting question opened by our study is whether typical degenerate and nondegenerate energy eigenstates at SUSY points exhibit distinct entanglement entropies.

Our other main finding is the polynomial decay in $L_A$ of the normalized standard deviation $\Sigma_S(\nu)\propto 1/\sqrt{L_A}$ in the eigenstates of the integrable spin-$\frac{1}{2}$ $XYZ$ and $XXZ$ chains. For the quantum-chaotic counterpart of those chains, we find an exponential decay in $L_A$. These results indicate that in the presence of interactions at integrability, even if the normalized standard deviation $\Sigma_S$ vanishes in the thermodynamic limit, the fluctuations of the eigenstate entanglement entropies are fundamentally different from (exponentially larger than) those in quantum-chaotic interacting Hamiltonians.

\acknowledgements
We acknowledge support from the Slovenian Research and Innovation Agency (ARIS), Research core fundings Grants No.~P1-0044 (R.S.~and L.V.), No.~N1-0273 and No.~J1-50005 (L.V.), the Polish National Agency for Academic Exchange (NAWA) Grant No.~PPI/APM/2019/1/00085 (M.K.), and the United States National Science Foundation (NSF) Grants No.~PHY-2012145 and~PHY-2309146 (M.R.). We acknowledge discussions with E. Ilievski, P. Łydżba, and M. Mierzejewski. 

\appendix

\section{Normalized standard deviations\label{app:entropy_structure}}

\begin{figure}[!t]
\includegraphics[width=0.94\columnwidth]{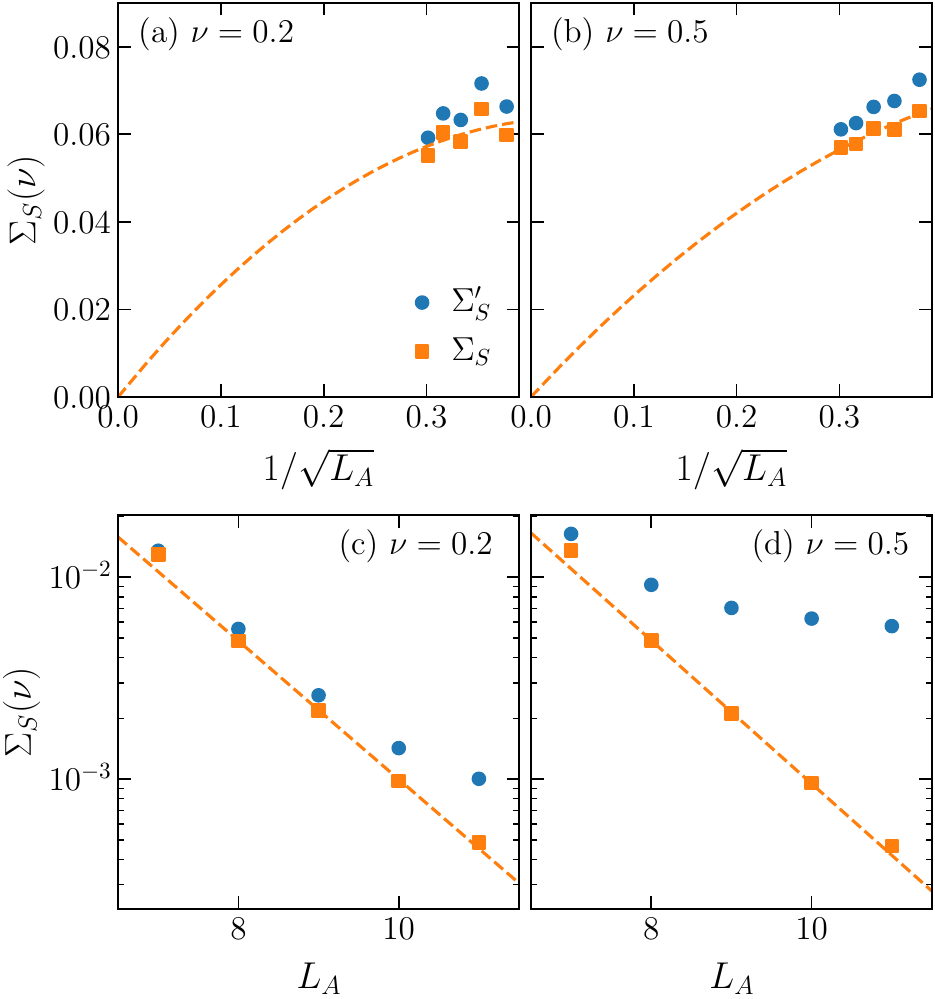}
\vspace{-0.15cm}
\caption{Scaling of the normalized standard deviations of the eigenstate entanglement entropy $\Sigma_{S}(\nu)$ [Eq.~\eqref{eq:entropy_var}] and $\Sigma'_{S}(\nu)$ [Eq.~\eqref{eq:entropy_var_trad}] for the $XYZ$ model with $\Delta=\eta=0.2$, corresponding to the result shown in Fig.~\ref{fig:entropy_structure}. [(a) and (b)] integrable interacting point, see Eq.~\eqref{eq:model:xyz}, and [(c) and (d)] quantum-chaotic interacting point (with $J_2=2$), see Eq.~\eqref{eq:model:xyzQC}. [(a) and (c)] $\nu=0.2$ and [(b) and (d)] $\nu=0.5$. The dashes lines show fits of $a/\sqrt{L_A}+b/L_A$ to the data for $\Sigma_{S}(\nu)$ in (a) and (b), and of $ae^{-bL_A}$ to the data for $\Sigma_{S}(\nu)$ in (c) and (d). For the latter fit, we use the largest four system sizes.}
\label{fig:entropy_structure_variance}
\end{figure}

Here, we compare results obtained for $\Sigma_{S}(\nu)$ as defined in Eq.~\eqref{eq:entropy_var} with results obtained using the traditional definition that does not account for the underlying structure of $S_{A}^\alpha$, namely,
\begin{equation}\label{eq:entropy_var_trad}
    \Sigma'_{S}(\nu)=\frac{\sqrt{\,\overline{(S_{A}^\alpha)^2}(\nu)-\bar S_{A}^{\,2}(\nu)}}{L_A\ln2},
\end{equation}

In Fig.~\ref{fig:entropy_structure_variance}, we show the scaling of $\Sigma_{S}(\nu)$ and $\Sigma'_{S}(\nu)$ for the eigenstate entanglement entropies reported in Fig.~\ref{fig:entropy_structure}. We show results at the integrable [nonintegrable] point for $\nu=0.2$ in Fig.~\ref{fig:entropy_structure_variance}(a) [Fig.~\ref{fig:entropy_structure_variance}(c)] and $\nu=0.5$ in Fig.~\ref{fig:entropy_structure_variance}(b) [Fig.~\ref{fig:entropy_structure_variance}(d)]. One can see that while $\Sigma_{S}(\nu)$ and $\Sigma'_{S}(\nu)$ exhibit qualitatively similar behavior at the integrable point [Figs.~\ref{fig:entropy_structure_variance}(a) and~\ref{fig:entropy_structure_variance}(b)], their behavior is qualitatively different at the quantum-chaotic point [Figs.~\ref{fig:entropy_structure_variance}(c) and~\ref{fig:entropy_structure_variance}(d)]. In Figs.~\ref{fig:entropy_structure_variance}(c) and~\ref{fig:entropy_structure_variance}(d), $\Sigma_{S}(\nu)$ decays exponentially with $L_A$ while $\Sigma'_{S}(\nu)$ decays slower than exponentially with $L_A$, and the larger the value of $\nu$ the slower the decay of $\Sigma'_{S}(\nu)$. Hence, if one does not subtract the underlying structure of $S_{A}^\alpha$ in the quantum-chaotic point, the exponential decay is not visible unless one carries out the averages in the limit $\nu\rightarrow 0$.

\newpage
\bibliographystyle{biblev}
\bibliography{references}

\begin{thebibliography}{10}
\bibitem{U1note}
\bibinfo{note}{The Hamiltonian for $\Delta=-(1\pm\eta)$ has the form $\hat
  H=\sum_\ell
  \frac{1\mp\eta}{2}\qty(\hat{s}^+_\ell\hat{s}^+_{\ell+1}+\hat{s}^-_\ell\hat{s}^-_{\ell+1})+\qty(1\pm\eta)\hat{s}^z_\ell\hat{s}^z_{\ell+1}$.
  It becomes the XXZ model after a parity transformation on the even sites like
  the one used to obtain Eq.~\eqref{eq:xyz:susy}.}
\end{thebibliography}

\end{document}